\documentclass[11pt]{article}
\usepackage[round]{natbib}
\usepackage{bbm}
\usepackage{bm}
\usepackage{graphicx}
\usepackage{amsmath}
\usepackage[utf8]{inputenc}
\usepackage[english]{babel}
 \usepackage[symbol]{footmisc}
\usepackage{todonotes}
\usepackage[export]{adjustbox}

\usepackage{wrapfig}
\usepackage{amsthm}
 \usepackage{amssymb} 
 \usepackage{subcaption}
 \usepackage{mwe}
 \usepackage{url}
 \usepackage{adjustbox}

\def\bse{\begin{eqnarray*}}
	\def\ese{\end{eqnarray*}}
\def\be{\begin{eqnarray}}
\def\ee{\end{eqnarray}}

\newtheorem{Assumption}{\underline{\bf Assumption}}
\newtheorem{Th}{\underline{\bf Theorem}}

\newtheorem{Cor}{\underline{\bf Corollary}}
\usepackage{ amsmath}
\usepackage{lineno}
\usepackage{bbm}
\usepackage{graphicx}
\usepackage{url}
\usepackage{multirow}

 \usepackage{setspace}

\makeatletter

\definecolor{maroon(html/css)}{rgb}{0.5, 0.0, 0.0}
\newcommand{\blind}{1}

\begin{document}

   \if1\blind
{
    \title{\bf A Semiparametric Quantile Single-Index Model for Zero-Inflated Outcomes}
  \author{Zirui Wang\\
    Department of Statistics and Data Science, Tsinghua University\\
    and \\
    Tianying Wang \\
    Department of Statistics, Colorado State University\\
    Tianying.Wang@colostate.edu}
  \maketitle

  \maketitle
} \fi

\if0\blind
{
  \bigskip
  \bigskip
  \bigskip
  \begin{center}
     {\LARGE\bf A Semiparametric Quantile Single-Index Model for Zero-Inflated Outcomes}
     
\end{center}
  \medskip
} \fi


\begin{abstract}
\noindent {\it Abstract:}
     We consider the complex data modeling problem motivated by the zero-inflated and overdispersed data from microbiome studies. Analyzing how microbiome abundance is associated with human biological features, such as BMI, is
of great importance for host health. Methods based on parametric distributional
assumptions, such as zero-inflated Poisson and zero-inflated Negative Binomial
regression, have been widely used in modeling such data, yet the parametric assumptions are restricted and hard to verify in real-world applications. We relax
the parametric assumptions and propose a semiparametric single-index quantile
regression model. It is flexible to include a wide range of possible association functions and adaptable to the various zero proportions across subjects, which relaxes
the strong parametric distributional assumptions of most existing zero-inflated
data modeling approaches. We establish the asymptotic properties for the index
coefficients estimator and quantile regression curve estimation. Through extensive simulation studies, we demonstrate the superior performance of the proposed
method regarding model fitting.
	\end{abstract}

\par

\def\thetable{\arabic{table}}

\renewcommand{\theequation}{\thesection.\arabic{equation}}

\fontsize{11}{14pt plus.8pt minus .6pt}\selectfont
\def\spacingset#1{\renewcommand{\baselinestretch}%
{#1}\small\normalsize} \spacingset{2}

\section{Introduction}

The human microbiota consists of the microorganisms that reside in or on the human body and 
contribute essential functions to human beings \citep{cani2018human}. Human microbiome 
research studies the dynamic interactions among microbiomes, host, and environment 
\citep{xia2017hypothesis}. It is of great importance to build more accurate predictive models of 
taxa and identify the relationship between taxa and clinical parameters \citep{lloyd2016healthy}. 
The main challenges in modeling microbiome data are zero inflation and 
overdispersion \citep{kaul2017analysis}. 
It is common that the proportion of zeros in gut microbiota counts can reach 70\%-80\% \citep{yatsunenko2012human}. Meanwhile, the non-zero counts of the microbiota counts could be as large as thousands and cause overdispersion \citep{mcmurdie2014waste}. The inflated zeros in microbiome data are commonly caused by two reasons: microbes are present in the environment but not detected due to low sequencing depth and sampling variation, or some microbes may be incapable of living in the environment and truly never represented \citep{zeng2022mbdenoise}. While modeling microbiota counts and testing their relationship with covariates of interest (e.g., lifestyle and disease status), one needs to carefully address the zero inflation and overdispersion challenges in statistical analysis \citep{zhang2017negative,xia2020correlation}.

 A common strategy to model zero-inflated data is two-part models, which impose a point probability mass at zero and model the positive count data by a parametric distribution, such as zero-inflated Poisson regression, zero-inflated Negative Binomial regression, hurdle models, and many others \citep{lambert1992zero,chen2016two,jiang2022flexible}. However, those approaches impose strong parametric assumptions, which may be violated in real-world applications and cause problems in downstream analysis \citep{silverman2020naught}. Further, most of the aforementioned approaches fail to model the relationship between the proportion of zeros and covariates, and thus, they only capture the effect of covariates partially on the distribution of outcomes \citep{ling2022statistical}.  

 Contrary to parametric modeling, quantile regression \citep{KoenkerBassett1978} is a powerful and robust tool to model heterogeneous associations in complex data without any parametric distribution assumptions. Further, quantile regression enjoys the merits of flexibly linking covariates to the distribution of response without parametric assumptions and allowing different associations across quantile levels.  However, classic quantile regression cannot be directly applied in microbiome data as it assumes a constant probability of observing a positive outcome for all individuals, which is unlikely to hold when the degree of zero inflation varies across subjects. To overcome this challenge, \cite{ling2022statistical} proposed a zero-inflated linear quantile regression model (denoted as ``ZIQ-linear")  to relax the parametric distribution assumptions on positive outcomes and applied this method in a carotid plaque data analysis. However, it does not consider either the overdispersion issue or the non-linear relationship between the quantile of microbiome data and the covariates of interest.

\begin{figure}[!ht]
    \centering\includegraphics[scale = 0.4]{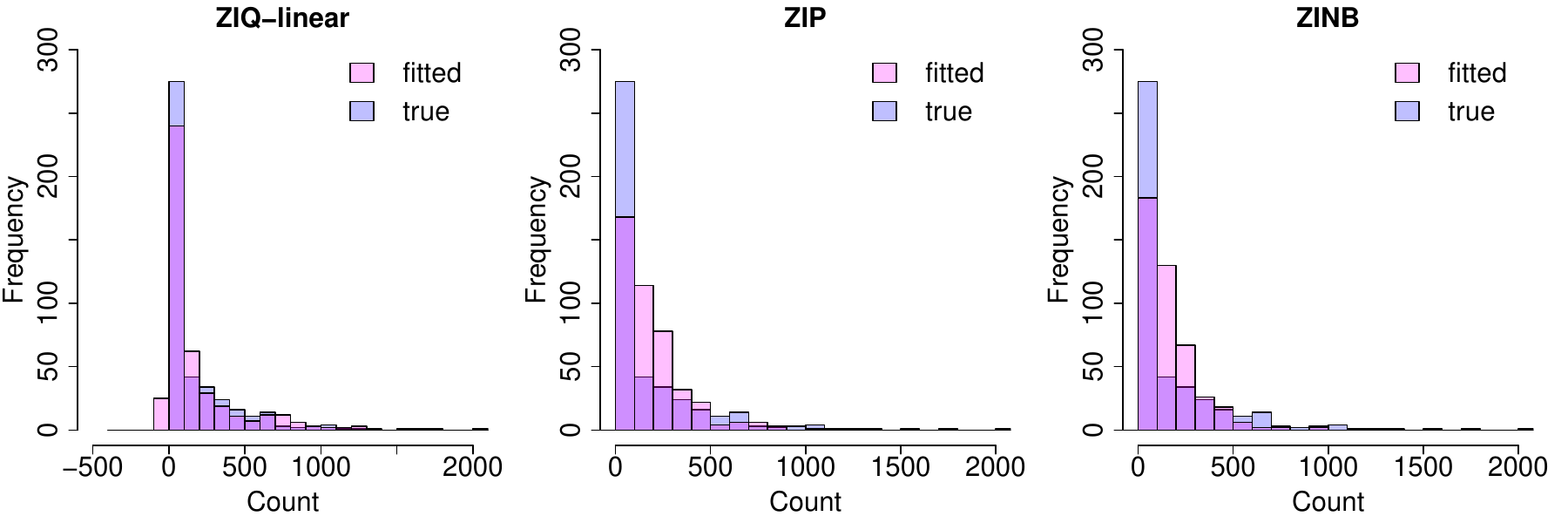}
 \caption{Model fitting results for the taxon $Clostridiales$.}\label{fig:hist_176062}
\end{figure}

We present a motivating example from the gut microbiota count data \citep{de2018gut}, which is later analyzed in the real data application. This dataset contains microbiome counts of over 6000 taxa for 411 adults and covariates related to diet, obesity, and cardiometabolic diseases. For illustrative purposes, we present the model-fitting results for one taxon $Clostridiales$ with health-related covariates, such as anthropometric measures, glucose
metabolism, and blood pressure. A full list of covariates is provided in Section \ref{sec:app}. The library size, the sum of all the taxa counts per subject, is adjusted as a covariate in the models below. We compared the observed and predicted count data generated by fitted models from ZIP, ZINB, and ZIQ-linear, respectively. In Figure \ref{fig:hist_176062}, we observe that ZIP and ZINB fit the data poorly because the parametric assumptions could be violated, and the mass probability imposed on zero is a shared parameter for all subjects in these two approaches rather than modeling different degrees of zeros across subjects. ZIQ-linear models the zero and positive parts well because these two parts are both linked to individual-specific covariates. However, a small proportion of fitted values are negative, against the nature of microbiome counts. The primary reason is that the linear quantile regression model is not flexible enough for overdispersed data. 

Motivated by the study of zero-inflated and overdispersed outcomes, we consider quantile single-index models to overcome the limitations of the linear quantile regression model while maintaining its robustness and flexibility. Single-index models have been widely used in literature for their merits of handling high-dimensional data while providing interpretable results \citep{radchenko2015high,neykov2016l1}. Spline-based methods are often preferred for their easy implementation and derivable asymptotic properties \citep{yu2002penalized,ma2013doubly}. 
For quantile regression, \cite{ma2016optimization} developed statistical inference for single-index quantile regression models based on the pseudo-profile likelihood approach, yet it cannot be directly applied to the zero-inflated data. Without two-part modeling, the direct quantile single-index models assume a constant chance of observing a positive outcome and ignore the various degrees of zero inflation across subjects.

To this end, we propose a novel two-part modeling approach: the Zero-Inflated Quantile Single-Index model (ZIQSI). The positive part $Y>0$ is modeled by a quantile single-index model in a semiparametric fashion, which is flexible and general, including a wide range of association functions. Compared to a fully nonparametric model, the proposed method is more interpretable through the index parameter. The probability of being zero (i.e., $P(Y=0)$) is also linked to covariates, making our approach more adaptable to various zeros across subjects.  Our contributions are three-fold. Methodologically, we provide a flexible modeling approach of zero-inflated and overdispersed outcomes with less restricted model specifications. The estimation is proceeded by the profile likelihood approach. Theoretically, we derived asymptotic properties for the estimated quantile coefficients, quantile curves, and average quantile effects.  Application-wise, we provided a concrete analysis of microbiome data and evaluated the goodness of fit for the proposed method from different perspectives, illustrating its superiority in both distribution-wise modeling and individual-wise coefficient estimation, which could further contribute to personalized medicine. 

\section{Methods }\label{sec:method}

\subsection{Notations and model}\label{sec:notation}

Denote $Y$ as a non-negative zero-inflated response variable and  $\textbf{x} = (x_1,...,x_p)^\top$ be a set of covariates of our interest. Denote the $\tau$th quantile of $Y$ as $Q_Y(\tau\mid\textbf{x})$.  To model the distribution of $Y$, we first decompose the conditional distribution of the zero-inflated outcome $Y$ into the zero part and the positive part:
$F(Y\mid\textbf{x}) = P(Y=0\mid\textbf{x})+P(Y>0\mid\textbf{x})F(Y\mid\textbf{x},Y>0).$
Then, following the common two-part modeling strategy,  we model the two parts, namely $P(Y=0\mid\textbf{x})$ and $F(Y\mid\textbf{x},Y>0)$, separately. We first assume that $P(Y>0\mid\textbf{x})$, the conditional probability of observing a positive $Y$, follows a logistic regression model,
\be\label{eq:logit}
{\rm logit}\left\{P(Y>0\mid\textbf{x})\right\} = \textbf{x}^\top\gamma,
\ee
where $\gamma$ is an unknown parameter. {We consider the linear form for the logistic regression model since no compelling evidence suggests that a complicated semiparametric model is necessary for the classification, whereas our motivating example indicates the crucial need to consider a more flexible model for the positive response $Y>0$.} {Thus, to ensure the generality of our method, we adopt a semi-parametric approach for the non-zero part $F(Y\mid\textbf{x},Y>0)$.} Given a nominal quantile level $\tau_s\in(0,1)$, the conditional quantile function of $Y$ given $Y>0$ can be described by a single-index model:
\be\label{single}
Q_Y(\tau_s\mid\textbf{x}, Y>0) = G_{\tau_s}(\textbf{x}^\top\beta_{\tau_s}),
\ee
where $G_{\tau_s}(\cdot)$ is an unknown function, and $\beta_{\tau_s}$ is an unknown parameter. The single-index model (eq \eqref{single}) is a popular dimensional reduction method for high-dimensional covariates $\textbf{x}$ with extra flexibility at each quantile level $\tau_s$ through the unknown function $G_{\tau_s}(\cdot)$, which is essential for modeling the overdispersion in microbiome data.  The method of \cite{ling2022statistical} can be viewed as a special case of our method, in which $G_{\tau_s}(\textbf{x}^\top\beta_{\tau_s})$ is set as $\textbf{x}^\top\beta_{\tau_s}$ for all $\tau_s \in (0,1)$. To ensure the continuity of this two-part model, we assume that for any $\textbf{x}$, $\lim_{\tau_s\to 0^+}Q_{Y}(\tau_s\mid\textbf{x},Y>0)= 0.$
Thus, when considering Models \eqref{eq:logit}-\eqref{single} together, the $\tau$th conditional quantile of $Y$ given $\textbf{x}$ can be written as:
\be\label{model:real}
Q_Y(\tau\mid\textbf{x}) = I\left\{\tau>1-\pi(\textbf{x},\gamma)\right\}G_{\tau_s}\left\{\textbf{x}^\top\beta_{\tau_s}\right\},
\ee
where $\pi(\textbf{x},\gamma) = P(Y>0\mid\textbf{x})$;  $I(\cdot)$ is an indicator function; and $\tau_s = \Gamma(\tau;\textbf{x},\gamma)=\max\left(\frac{\tau-\left\{1-\pi(\gamma,\textbf{x})\right\}}{\pi(\gamma,\textbf{x})},0\right)$ maps the target quantile level $\tau$ linearly to the quantile level $\tau_s$ of $Y\mid Y>0$ in Model \eqref{single}.
Due to the nonparametric nature of $G_{\tau_s}$, we posit the assumptions for model identifiability.
\begin{Assumption}\label{assum:id}
    \item[(1.1)] The covariates $\textbf{x}$ satisfies that  $\textbf{x}\in \mathcal{C}$, where $\mathcal{C}$ is a compact set.

\item[(1.2)] $\beta_{\tau_s}$ belongs to the parameter space $\Theta = \left\{\beta:\beta\in \mathbb{R}^p,\right\|\beta\|_2=1, \beta_1\geq 0\}$ for identifiability. We assume $p$ is fixed and shall not increase with $n$.
\item[(1.3)] Support of the function $G_{\tau_s}$ is $\left[\inf(\textbf{x}^\top\beta),\sup(\textbf{x}^\top\beta)\right]$, $\forall\textbf{x}\in \mathcal{C},\beta \in\Theta$. 
\end{Assumption}
These assumptions guarantee the identifiability of $\beta_{\tau_s}$ for quantile single-index models  \citep{ma2016optimization}. Compared to \cite{ling2022statistical}, our proposed two-part Zero-Inflated Quantile Single-index model allows more complex nonlinear associations between $\textbf{x}$ and $Y$ through the functions  $G_{\tau_s}$. Compared to other parametric two-part models, such as ZIP and ZINB, it is robust against non-gaussian errors because we do not assume any particular error distributions.

\subsection{Estimation}
Suppose we have independent and identically distributed random samples $\{(\textbf{x}_i,y_i);i = 1,2,...,n\}$ generated by the conditional quantile regression model \eqref{model:real}. First, we estimate $\gamma$ by logistic regression model \eqref{eq:logit}: 
\[\hat\gamma_n = \arg\max_{\gamma}\frac{1}{n}\sum_{i=1}^n\left[I(y_i>0)\log\left\{\frac{\pi(\gamma,\textbf{x}_i)}{1-\pi(\gamma,\textbf{x}_i)}\right\}+\log\{1-\pi(\gamma,\textbf{x}_i)\}\right].\]
With the estimated coefficient $\hat\gamma_n$, given $\textbf{x},\tau$, we approximate $\tau_s$ by 
\be\label{eq:taus}
\hat\tau_s = \Gamma(\tau;\textbf{x},\hat\gamma_n) = \max\left(\frac{\tau-\left(1-\pi(\hat\gamma_n,\textbf{x})\right)}{\pi(\hat\gamma_n,\textbf{x})},0\right).
\ee

For the quantile regression part, since $G_{\tau_s}(\cdot)$ is unknown, we approximate it by a linear combination of B-spline basis functions as in \cite{wei2006conditional}. We first introduce the B-spline basis for estimating the unknown function $G_{\tau_s}$. Denote the total number of positive responses as $n_0 := \sum_{i=1}^n I(y_i>0)$. We denote $a=t_0<t_1<...<t_{N_{n_0}}<b=t_{N_{n_0}+1}$ as a partition of $[a,b]$, where the number of knots $N_{n_0}$ increases with $n_0$. The partition satisfies
$\max_{0\le j\le N_{n_0}}|t_{j+1}-t_j|/\min_{0\le j\le N_{n_0}}|t_{j+1}-t_j|\le M$
uniformly in the sample size of positive outcomes ${n_0}$ and for some constant $0<M<\infty$. With $m$ denoted as the order of polynomial splines, we denote the normalized B-spline basis of this space \citep{de2001revised}, as $B(u) = \{B_j(u):1\le j\le J_n\}^\top$, where $J_n=N_{n_0}+m$. In our empirical implementations, for each given $\beta$, we use the boundary points, namely $\min_{1\le i\le {n}}\textbf{x}_i^\top\beta$ and $\max_{1\le i\le n}\textbf{x}_i^\top\beta$, to generate the B-spline basis function $B(u)$. Further, by \cite{de2001revised}, the single-index term $G_{\tau_s}(\textbf{x}^\top\beta_{\tau_s})$ can be approximated by B-spline as $G_{\tau_s}\left(\textbf{x}^\top\beta_{\tau_s}\right)\approx B\left(\textbf{x}^\top\beta_{\tau_s}\right)^\top\theta(\tau_s)$ for some $\theta(\tau_s)\in \mathbb{R}^{J_n}$. 
Since the true value of $\tau_s$ is infeasible, we  use its approximation $\hat\tau_s$ defined in eq \eqref{eq:taus}  to obtain the estimators of the spline coefficients $\theta(\tau_s)$ and the parameter  $\beta_{\tau_s}$ by minimizing the pseudo-likelihood function:
\be\label{eq:loss_function}
L_{\hat\tau_s, n}(\theta,\beta) &=&\frac{1}{n_0}\sum_{i=1}^n\rho_{\hat\tau_s}\left\{y_i-B(\textbf{x}_i^\top\beta)^\top\theta\right\}I(y_i>0),
\ee
where $\rho_\tau(u) = u\left(\tau-I(u<0)\right)$ is the quantile loss function.

Here, we adopt the profile approach proposed in \cite{ma2016optimization} to estimate $\beta_{\tau_s}$ and $\theta(\tau_s)$ owing to the stable performance showed in the empirical studies of \cite{liang2010estimation} and \cite{ma2016optimization}. We define the profile pseudo-likelihood function of $\beta$ as
\be\label{eq:loss_beta}
L_{\hat\tau_s, n}^*(\beta)&=&\min_{\theta\in\mathbb{R}^{J_n}} L_{\hat\tau_s, n}(\beta,\theta)= L_{\hat\tau_s, n}\left(\beta,\tilde\theta_n(\beta,\hat\tau_s)\right)\nonumber\\&=&\frac{1}{n_0}\sum_{i=1}^n\rho_{\hat\tau_s}\left\{y_i-B(\textbf{x}_i^\top\beta)^\top\tilde\theta_n(\beta,\hat\tau_s)\right\}I(y_i>0),
\ee
where $\tilde\theta_n(\beta,\hat\tau_s)$ is the minimizer of $L_{\hat\tau_s}(\theta,
\beta)$ over $\theta\in\mathbb{R}^{J_n}$ for given $\beta\in\Theta$.
Thus, the proposed profile likelihood estimation of $ \beta\circ\Gamma(\tau;\textbf{x},\hat\gamma_n)$ is taken to be:
\[\hat\beta_{\hat\tau_s} =\hat\beta\circ\Gamma(\tau;\textbf{x},\hat\gamma_n) =  \arg\min_{\beta\in\Theta}L_{\hat\tau_s, n}^*(\beta).\]
Then, the spline estimator of $G_{\tau_s}(u)$ is  $\widehat{G}_{\hat\tau_s }\left(u,\hat\beta_{\hat\tau_s}\right) = B(u)^\top\tilde\theta_n\left(\hat\beta_{\hat\tau_s},\hat\tau_s\right)$, where 
$\tilde\theta_n\left(\hat\beta_{\hat\tau_s},\hat\tau_s\right)$ minimizes $L_{\hat\tau_s, n}^{**}(\theta)$ over $\theta\in\mathbb{R}^{J_n}$, and
$
L_{\hat\tau_s, n}^{**}(\theta)= \frac{1}{n_0}\sum_{i=1}^n\\\rho_{\hat\tau_s}\left\{y_i-B(\textbf{x}_i^\top\hat\beta_{\hat\tau_s})^\top\theta\right\}I(y_i>0).
$

For a given $\beta\in\Theta$ and a specific $\tau_s$, we denote: 
\be\label{eq:tilde2}
\tilde{\tilde{\theta}}_{n}(\beta,\tau_s) = \arg\min_{\theta\in\mathbb{R}^{J_n}}\mathbb{E}\{L_{\tau_s, n}(\theta,\beta)\mid\mathbb{X}\},
\ee
where $L_{\tau_s, n}(\theta,\beta)$ is the score function eq \eqref{eq:loss_function} and $\mathbb{X}$ are given covariates.  We denote $\tilde{\tilde{G}}_{\tau_s}(u,\beta) = B^\top(u)\tilde{\tilde{\theta}}_{n}(\beta,\tau_s)$, which bridges the estimated $\widehat{G}_{\hat\tau_s}(\textbf{x}^\top\hat\beta_{\hat\tau_s})$ and true $G_{\tau_s}(\textbf{x}^\top\beta_{\tau_s})$. We also define 
\be\label{eq:tildex}
E^*(\textbf{x}\mid \textbf{x}^\top\beta_{\tau_s}) = \frac{\mathbb{E}\{f_{\epsilon_{\tau_s}}(0\mid\textbf{x})\textbf{x}\mid \textbf{x}^\top \beta_{\tau_s}\}}{\mathbb{E}\{f_{\epsilon_{\tau_s}}(0\mid\textbf{x})\mid\textbf{x}^\top \beta_{\tau_s}\}} \ \ \text{and}\ \ \tilde{\textbf{x}} = \textbf{x} - E^*(\textbf{x}\mid \textbf{x}^\top\beta_{\tau_s}),
\ee
where $f_{\epsilon_{\tau_s}}(\epsilon\mid\textbf{x})$ denotes the conditional density of $\epsilon_{\tau_s}$ given $\textbf{x}$, and $\epsilon_{\tau_s} = Y - G_{\tau_s}(\textbf{x}^\top \beta_{\tau_s})$ given $Y>0$. $E^*(\textbf{x}\mid \textbf{x}^\top\beta_{\tau_s})$ and $\tilde{\textbf{x}}$ are necessary for deducing the asymptotic distribution for the estimated coefficient $\hat\beta_{\hat\tau_s}$.
 
\subsection{Construction of Quantile Curve and Average Quantile Effect}\label{sec:quantile curve} 

Given the aforementioned estimators $\hat\gamma_n$, $\hat\beta_{\hat\tau_s}$, and $\tilde{\theta}_n(\hat\beta_{\hat\tau_s},\hat\tau_s)$, we construct the $\tau$th conditional quantile function $\widehat{Q}_Y(\tau\mid\textbf{x})$ in three regions:
(1) $R_{1,n} = \left\{\tau:0<\tau<1-\pi(\hat\gamma_n,\textbf{x})\right\}$, (2) $R_{2,n} = \left\{\tau:1-\pi(\hat\gamma_n,\textbf{x})\le\tau\le1-\pi(\hat\gamma_n,\textbf{x})+n^{-\delta}\right\}$, and (3) $R_{3,n} = \left\{1-\pi(\hat\gamma_n,\textbf{x})+n^{-\delta}\le\tau\le 1\right\},
$
where $\delta<0.5$ is a pre-specified interpolation parameter, and 
$\pi(\hat\gamma_n,\textbf{x}) = 
\exp(\textbf{x}^\top\hat\gamma_n)/\{1+\exp(\textbf{x}^\top\hat\gamma_n)\}$ is the 
estimated probability of observing a positive $Y$ given $\textbf{x}$. Specifically, $R_{1,n}$ represents the region for a zero $Y$; $R_{3,n}$ represents the region of the positive part, in which the quantile curve is estimated on the nominal quantile level $\hat\tau_s = \Gamma(\tau;\textbf{x},\hat\gamma_n)$. $R_{2,n}$ is an interpolation region based on the nominal quantile level $\Gamma\left(1-\pi(\hat\gamma_n,\textbf{x})+n^{-\delta};\textbf{x},\gamma\right)$. The conditional density of $Y$ given $Y>0$ goes to zero when the quantile level approaches the change point, which can lead to a large variance if we estimate the quantile directly around the change point. The interpolation region is set for the stability and continuity of the estimated quantile function  $\widehat{Q}_Y(\tau\mid\textbf{x})$. Then, we construct  $\widehat{Q}_Y(\tau\mid\textbf{x})$  as below:
\be\label{eq:estimated_quantile}
\widehat{Q}_Y(\tau\mid\textbf{x})
&=& 0\cdot I(\tau\in R_{1,n})
+ B\left\{\textbf{x}^\top\hat\beta\circ\Gamma\left(1-\pi(\hat\gamma_n,\textbf{x})+n^{-\delta};\textbf{x},\hat\gamma_n\right)\right\}^\top\nonumber\\
&&\tilde\theta_n\left\{\hat\beta\circ\Gamma\left(1-\pi(\hat\gamma_n,\textbf{x})+n^{-\delta};\textbf{x},\hat\gamma_n\right),\Gamma\left(1-\pi(\hat\gamma_n,\textbf{x})+n^{-\delta};\textbf{x},\hat\gamma_n\right)\right\}\nonumber\\
&&\cdot\frac{\tau-\{1-\pi(\hat\gamma_n,\textbf{x})\}}{n^{-\delta}}\cdot I(\tau\in R_{2,n})+ B\left\{\textbf{x}^\top\hat\beta\circ\Gamma(\tau;\textbf{x},\hat\gamma_n)\right\}^\top\nonumber\\
&&\cdot\tilde\theta_n\{\hat\beta\circ\Gamma(\tau;\textbf{x},\hat\gamma_n),\Gamma(\tau;\textbf{x},\hat\gamma_n)\}\cdot I(\tau\in R_{3,n}).
\ee
 Based on $\widehat{Q}_Y(\tau \mid \textbf{x})$, it is obvious that the covariates $\textbf{x}$ can be associated with both the probability of observing a positive $Y$ and also the quantile of $Y\mid Y>0$. { As our main focus is predicting quantile curves, it is common that the predicted values are non-integral. One can round the estimation to the nearest integer upon request. The same applies to the following data simulation settings in Section \ref{sec:sim_set}.}

To quantify the effect of the $j$th covariate (denoted as $x_j$) on the response $Y$, we define the model-based average quantile effect (AQE) for our two-part model as below:
\be\label{eq:define_AQE}
\Delta_{\tau}(x_j;u,v) = \mathbb{E}_{\textbf{x}^{(-j)}}\left\{Q_Y(\tau\mid x_j=u,\textbf{x}^{(-j)})-Q_Y(\tau\mid x_j=v,\textbf{x}^{(-j)})\right\},
\ee
where  $\textbf{x}^{(-j)}$ denotes the covariates excluding $x_j$. AQE is served in an analogous fashion to the average treatment effect in linear models, and it has also been used in \cite{ling2022statistical}. Thus, at a fixed quantile level $\tau$, the importance of the covariate $x_j$ can be estimated by integrating the difference between the conditional quantile of $Y$, given fixed $\textbf{x}^{(-j)}$ and different levels of $x_j$. If $x_j$ represents a continuous variable (e.g., BMI, cholesterol), we may select two levels according to clinical interest as $u$ and $v$. For example, to assess the quantile effect of BMI, one can set $u\in[18.5,24.9]$ for the normal group and $v \in[25,29.9]$ for the overweight group \citep{weir2019BMI}. In particular, if $x_j$ is binary (e.g., sex, treatment), the AQE is the average quantile treatment effect in the source population:
\be\label{eq:define_AQSE}
\Delta_{\tau}(x_j;1,0) = \mathbb{E}_{\textbf{x}^{(-j)}}\left\{Q_Y\left(\tau\mid x_j=1,\textbf{x}^{(-j)}\right)-Q_Y\left(\tau\mid x_j=0,\textbf{x}^{(-j)}\right)\right\}.
\ee
A natural sample estimator of eq \eqref{eq:define_AQSE} is
\be\label{eq:estimate_AQE}
\widehat\Delta_{\tau}(x_j;u,v) = \frac{1}{n}\sum_{i=1}^n\widehat{Q}_Y(\tau\mid x_{i,j}=1,\textbf{x}_i^{(-j)})-\widehat{Q}_Y(\tau\mid x_{i,j}=0,\textbf{x}_i^{(-j)}), 
\ee
where $\widehat{Q}_Y(\cdot)$  is the estimated conditional quantile function defined in eq \eqref{eq:estimated_quantile} and $(x_{i,j},\textbf{x}_i^{(-j)})$ denote the corresponding covariates of the $i$th sample.
 We provide the convergence rate of the AQE in Supplement S1.2.

\subsection{Assumptions for asymptotic properties}\label{sup:assum}
We introduce some common assumptions on the distribution of zero-inflated data. We denote $a_0$ and $b_0$ to be the infimum and supremum of $\textbf{x}^\top\beta_{\tau_s}$ over $\textbf{x}\in\mathcal{C}$, where $\mathcal{C}$ is the compact set defined in Assumption \ref{assum:id} above. 
\begin{Assumption}\label{assum:asy}

    \item[(2.1)]\label{assum:2.1}
Observations $\{(\textbf{x}_i, y_i);i = 1,\cdots , n\}$ are i.i.d. from a joint distribution $P$, where $\textbf{x}_i$ is a p-dimensional vector of covariates.   
\item[(2.2)]\label{assum:2.2} The conditional density $f_Y(Y\mid\textbf{x},Y>0)$ of $Y$ given $X=\textbf{x}$ and $Y>0$ satisfies the Lipschitz condition of order $1$ and  $\sup_{\textbf{x},y}f_Y(Y\mid\textbf{x},Y>0)<\infty$.
\item[(2.3)]\label{assum:2.3} The conditional quantile function satisfies 
$\lim\limits_{\tau\to 0^+} Q_Y(\tau\mid\textbf{x},Y>0) = 0.$
\item[(2.4)]\label{assum:2.4} The quantile coefficient $\beta_{\tau_s}$ is a differentiable function of $\tau_s$ with bounded first derivative, i.e.,
$\sup_{\tau_s\in(0,1)}\dot\beta_{\tau_s} = \sup_{\tau_s\in(0,1)}\left.\dfrac{d\beta_t}{dt}\right|_{t = \tau_s}<\infty$.
\item[(2.5)]$\forall \textbf{x}\in\mathcal{C}$, we have $\|E(\textbf{x}\textbf{x}^\top)\|_{\infty}<\infty$.
\item[(2.6)]The density function of $\textbf{x}^\top\beta$ is bounded away from zero and infinity on its support, for $\beta$ in a neighborhood of $\beta_{\tau_s}$.
\end{Assumption}

Assumption $(2.2)$ is borrowed from \cite{ma2016optimization} to help establish the limiting distribution at the change point $\tau = 1-\pi(\gamma, \textbf{x})$. Assumption $(2.3)$ is the continuity assumption stated in Section \ref{sec:notation}. Assumptions $(2.4)-(2.5)$ and Assumption \ref{assum:Gtau} below are necessary for establishing the asymptotic distribution of estimated coefficient $\hat\beta_{\hat\tau_s}$. With the asymptotic distribution of estimated coefficient $\hat\beta_{\hat\tau_s}$, Assumption $(2.6)$ and Assumption \ref{assum:Gtau}, we can provide the convergence rate of $\widehat{Q}_Y(\tau\mid\textbf{x})$ for any $\tau>1-\pi(\gamma,\textbf{x})$.
The limiting distribution of $\widehat{Q}_Y(\tau\mid\textbf{x})$ at the change point $\tau = 1-\pi(\gamma,\textbf{x})$ is then proved based on the assumptions above and the asymptotic properties of $\widehat{Q}_Y(\tau\mid\textbf{x})$ when $\tau>1-\pi(\gamma,\textbf{x})$.  

For $\widehat{Q}_Y(\tau\mid \textbf{x})$ given $\tau>1-\pi(\gamma,\textbf{x})$, since our proof concerns nonparametric smoothing
literature, we first give some definitions and notations. Let $\mathcal{H}_r$ be the collection of all the functions on $[a_0,b_0]$ such that the $m$th order derivative satisfies the H\"older condition of order $r-m$, i.e. for each function $\phi\in\mathcal{H}_r$, there exists a constant $C_0$ s.t. $\left|\phi^{(m)}(u_1)-\phi^{(m)}(u_2 )\right|\le C_0|u_1-u_2|^{r-m},$
for any $u_1,u_2\in [a_0,b_0]$. This collection of functions is essential for proving the convergence rate of the spline estimator of $G_{\tau_s}$. 

For given $\beta\in\Theta$ and $\tau$, we denote: $\tilde{\tilde{\theta}}_{n}(\beta,\tau) = \arg\min_{\theta\in\mathbb{R}^{J_n}}\mathbb{E}\{L_{\tau, n}(\theta,\beta)\mid\mathbb{X}\},$
where $L_{\tau, n}(\theta,\beta)$ is the score function eq \eqref{eq:loss_function} and $\mathbb{X}$ are given covariates whose  corresponding $Y>0$. We denote $\tilde{\tilde{G}}_{\tau, n}(u,\beta) = B^\top(u)\tilde{\tilde{\theta}}_{n}(\beta,\tau)$, where $\tilde{\tilde{\theta}}_{n}(\beta,\tau)$ is from eq \eqref{eq:tilde2}. Now we present assumptions for $\tilde{\tilde\theta}_n$ and $G_{\tau_s}(\cdot)$. 
\begin{Assumption}\label{assum:Gtau}

  \item[(3.1)] There exists $r>\frac{3}{2}$, such that for any $\tau_s\in(0,1)$  we have $G_{\tau_s}\in \mathcal{H}_r$.

\item[(3.2)] There exists a constant $c_0\in (0,+\infty)$, such that
\[\sup\limits_{\mathbb{X}}\left\|\partial\tilde{\tilde{G}}_{\tau_s, n}(\textbf{x}_i^\top\beta,\beta)/\partial\beta-\partial\tilde{\tilde{G}}_{\tau_s, n}(\textbf{x}_i^\top\beta_{\tau_s},\beta_{\tau_s})/\partial\beta\right\|_2\le c_0\|\beta-\beta_{\tau_s}\|_2.\]
for any $\beta$ in the neighborhood of $\beta_{\tau_s}$ and $\tau_s \in (0,1)$.
\item[(3.3)] For fixed $\textbf{x}$, assume $G_{\tau_s}(\textbf{x}^\top\beta_{\tau_s})$ has limited first order derivative with respect to $\tau_s$, i.e., $\sup_{\tau_s\in(0,1)}\left|\frac{\partial G_{\tau_s}(\textbf{x}^\top\beta_{\tau_s})}{\partial \tau_s}\right|<\infty$.

\item[(3.4)]
For any $\tau_s\in (0,1)$, $E^*(\textbf{x}\mid \textbf{x}^\top\beta_{\tau_s} = u)$, which is a function of $u$, has a continuous and bounded first derivative.

\end{Assumption}

Assumption $(3.1)$ is commonly used in the nonparametric smoothing
literature \citep{ma2016optimization}. Assumption $(3.2)$ is a typical assumption in the regression literature, which can
be easily satisfied when the dimension of covariates is fixed. Assumption $(3.3)$ is used in the proof of the convergence rate for $\tau>1-\pi(\textbf{x};\gamma)$. Assumption $(3.1)$-$(3.2)$ with Assumption \ref{assum:asy} provides the constraints of the asymptotic property of normal spline estimator provided in \cite{ma2016optimization}. Also, Assumption \ref{assum:Gtau} together with Assumption $(2.4)$-$(2.5)$ ensures that the following matrices exist and positive definite:
\begin{eqnarray*}
\Lambda_{1,\tau_s} &=& \mathbb{E}[\pi(\gamma,\textbf{x})f_{\epsilon_{\tau_s}}\{G_{\tau_s}(\textbf{x}^\top\beta_{\tau_s})\mid\textbf{x}\}\{G_{\tau_s}^{(1)}(\textbf{x}^\top\beta_{\tau_s})\tilde{\textbf{x}}\}^{\otimes 2}],\\
\Omega_{\tau_s} &=& \mathbb{E}[\{G_{\tau_s}^{(1)}(\textbf{x}^\top\beta_{\tau_s})\tilde{\textbf{x}}\}^{\otimes 2}], \quad
D_{1,\gamma} = \mathbb{E}[\pi(\gamma,\textbf{x})\{1-\pi(\gamma,\textbf{x})\}\textbf{x}\textbf{x}^\top],
\end{eqnarray*}
where $\tilde{\textbf{x}} = \textbf{x}-E^*(\textbf{x}\mid \textbf{x}^\top\beta_{\tau_s}),$ and  $A^{\otimes 2} = AA^\top$, and $G_{\tau_s}^{(1)}(\cdot)$ is the first derivative of $G_{\tau_s}(\cdot)$. Here the matrices $\Omega_{\tau_s}$ and $\Lambda_{1,\tau_s}$ are constructed to approximate the variance-covariance matrix for the parameter $\hat\beta_{\hat\tau_s}.$

\subsection{Asymptotic properties for estimation}\label{sup:thm}
 First,  we provide the asymptotic normality for the individual estimated single-index coefficient $\hat\beta\circ\Gamma(\tau;\textbf{x},\hat\gamma_n)$ using the property of B-spline estimator in \cite{ma2016optimization} and the property of logistic regression. Denote the Moore-Penrose inverse of a matrix $A$ as $A^+$. 

\begin{Th}\label{thm:normal} Suppose $n\to\infty$ and $n_0/n\to b_0$ with $0<b_0<1$. Under the Assumptions \ref{assum:id}-\ref{assum:Gtau}, for all $ \textbf{x}\in\mathcal{C}$, when $\tau>1-\pi(\gamma,\textbf{x})$, we have:
\[\sqrt{n}\left\{\hat\beta_{\hat\tau_s}-\beta_{\tau_s}\right\}=\sqrt{n}\left\{\hat\beta\circ\Gamma(\tau;\textbf{x},\hat\gamma_n)-\beta\circ\Gamma(\tau;\textbf{x},\gamma)\right\}\overset{d}{\to} N(0,\Sigma_1+\Sigma_2),\]
where 
\begin{eqnarray*}
\Sigma_1 &=& 
b_0^{-1/2}\Gamma(\tau;\textbf{x},\gamma)\left\{1-\Gamma(\tau;\textbf{x},\gamma)\right\}\Lambda_{1,\Gamma(\tau;\textbf{x},\gamma)}^+\Omega_{\Gamma(\tau;\textbf{x},\gamma)}\Lambda_{1,\Gamma(\tau;\textbf{x},\gamma)}^+,\\
\Sigma_2 &=& 
 \left\{1-\Gamma(\tau;\textbf{x})\right\}^2\left\{1-\pi(\gamma,\textbf{x})\right\}^2\textbf{x}^\top
  D_{1,\gamma}^{-1}\textbf{x} 
 \textbf{x}^\top\dot\beta\circ\Gamma(\tau;\textbf{x},\gamma)\dot\beta\circ\Gamma(\tau;\textbf{x},\gamma)^\top
  \textbf{x},\\
  \dot\beta\circ\Gamma(\tau;\textbf{x},\gamma) &=& \left.\dfrac{d\beta_{\tau}}{d\tau}\right|_{\Gamma(\tau;\textbf{x},\gamma).}
\end{eqnarray*}
\end{Th}

{The covariance matrices, $\Sigma_1$ and $\Sigma_2$, are constructed using B-spline-based single-index quantile regression and logistic regression, respectively, and are then combined through the delta method. Both $\Lambda_{1,\tau_s}$ and $\Omega_{\tau_s}$ are evaluated conditional on $Y>0$ and adjusted for the individual zero-inflation rate, $\pi(\gamma,\textbf{x})$. That is, $\pi(\gamma,\textbf{x})$ can be viewed as the propensity score to adjust for the covariance matrix since only the positive $Y$'s are considered to fit the quantile regression model.} 
Then, we construct the asymptotic consistency for $\widehat{Q}_Y(\tau\mid \textbf{x})$ in Theorem \ref{thm:consistency}.
\begin{Th}\label{thm:consistency} 
Suppose $n\to\infty$ and $n_0/n\to b_0$ with $0<b_0<1$. Under the Assumptions \ref{assum:id}-\ref{assum:Gtau}, we have
$\widehat Q_Y(\tau\mid\textbf{x})\overset{p}{\to} Q_Y(\tau\mid\textbf{x}).$
\end{Th}

Next, we provide the asymptotic properties for the limiting distribution of $\widehat{Q}_Y(\tau\mid \textbf{x})$ in Theorem \ref{thm:convergence}. For $\tau<1-\pi(\gamma, \textbf{x})$, $\widehat{Q}_Y(\tau\mid \textbf{x})$  converges to $0$ super-efficiently due to the property of logistic regression. For the change point  $\tau = 1-\pi(\gamma,\textbf{x})$, $\widehat{Q}_Y(\tau\mid \textbf{x})$ has different convergence conditions based on the parameter $\delta$. When $\tau > 1-\pi(\gamma,\textbf{x})$, the usage of the B-spline basis function makes it infeasible to establish the asymptotic distribution for $\widehat{Q}_Y(\tau\mid Y>0, \textbf{x})$ as the number of knots $N_{n_0}$ increases with $n_0$. Thus, we provide the global convergence rate for $\widehat{Q}_Y(\tau\mid Y>0, \textbf{x})$.

\begin{Th}\label{thm:convergence}  Under the conditions of Theorem \ref{thm:normal}-\ref{thm:consistency}, given $\textbf{x}$ and $\tau$, we have the asymptotic convergence  for the estimated quantile function as follows:
\item[(i)] when $\tau < 1-\pi(\gamma,\textbf{x})$, we have
$\sqrt n(\widehat Q_Y(\tau\mid\textbf{x})-0)\overset{p}{\to}  0;$
\item[(ii)] when $\tau = 1-\pi(\gamma,\textbf{x})$, we denote $Q_Y^{'}(0\mid\textbf{x},Y>0)$ as the right derivative and $Z_0\sim N(0,1)$, then we have:
\begin{enumerate}
    \item[(a)] when $\delta = 0.25$,
\[\sqrt{n}\{\widehat{Q}_Y(\tau\mid Y>0,\textbf{x})-0\}\overset{d}{\to}\{1-\pi(\gamma,\textbf{x})\}\sqrt{\textbf{x}^\top D_{1,\gamma}^{-1}\textbf{x}}\ Q_Y^{'}(0\mid\textbf{x},Y>0)Z_0I(Z_0>0);\]
\item[(b)] when $0.25<\delta < 0.5$, $\widehat{Q}_Y(\tau\mid Y>0,\textbf{x})-0 = O_P\left(J_n^{\frac{1}{2}}n^{-\frac{1}{2}}+J_n^{-r}\right);$

\end{enumerate}

\item[(iii)] when $\tau > 1-\pi(\gamma,\textbf{x})$, we have the global optimal convergence rate as\[\widehat{Q}_Y(\tau\mid Y>0,\textbf{x})-Q_Y(\tau\mid Y>0,\textbf{x}) = O_P\left(J_n^{\frac{1}{2}}n^{-\frac{1}{2}}+J_n^{-r}\right),\]i.e., $B\left(\textbf{x}^\top\hat\beta_{\hat\tau_s}\right)^\top\tilde\theta_n\left(\hat\beta_{\hat\tau_s},\hat\tau_s\right)-G_{\Gamma(\tau;\textbf{x},\gamma)}\left(\textbf{x}^\top\beta\circ\Gamma(\tau;\textbf{x},\gamma)\right)= O_P\left(J_n^{\frac{1}{2}}n^{-\frac{1}{2}}+J_n^{-r}\right),$ where $r$ is defined in Assumption $(3.1)$.

\end{Th}

{The asymptotic property at the change point mainly depends on the interpolation region $R_{2,n}$ with length $n^{-\delta}$, in which the threshold for $\delta$ is determined based on the convergence condition at $\tau = 1-\pi(\gamma,\textbf{x})$. When $\delta \leq 0.25$, the variance from quantile regression at the change point is controlled by $n^\delta$, allowing $\sqrt{n}$ convergence, but the slow convergence of the interpolation region leads to noticeable bias. For $\delta \in (0.25, 0.5)$, we achieve faster convergence of the interpolation region while keeping variance within a reasonable range. When $\delta \geq 0.5$, similar to the proof of Theorem \ref{thm:convergence} (ii)(b), the convergence rate at the change point slows, and larger $\delta$ values result in growing variance and unstable estimates. }In numerical studies, we set $\delta = 0.499$ as in \cite{ling2022statistical} for a fair comparison. We also provide results with $\delta = 0.250$ in Supplement S2.5 and S3.1, which suggests that the choice of $\delta$ does not affect the estimation results very much. 

From Theorem \ref{thm:convergence} $(iii)$, we have the following corollary directly.
\begin{Cor}\label{cor1}
When $\tau>1-\pi(\gamma,\textbf{x})$, under the conditions of Theorem \ref{thm:convergence}, we have
$\frac{1}{n}\sum_{i=1}^n\widehat{Q}(\tau\mid\textbf{x}_i)-Q(\tau\mid\textbf{x}_i) = O_P\left(J_n^{\frac{1}{2}}n^{-\frac{1}{2}}+J_n^{-r}\right),$
i.e.,
\[\frac{1}{n}\sum_{i=1}^n B\left(\textbf{x}_i^\top\hat\beta_{\hat\tau_s}\right)^\top\tilde\theta_n\left(\hat\beta_{\hat\tau_s},\hat\tau_s\right)-G_{\Gamma(\tau;\textbf{x}_i,\gamma)}\left(\textbf{x}_i^\top\beta\circ\Gamma(\tau;\textbf{x}_i,\gamma)\right)= O_P\left(J_n^{\frac{1}{2}}n^{-\frac{1}{2}}+J_n^{-r}\right).\]
\end{Cor}
The proofs for the theorems above are provided in Supplement S1.1.

\subsection{Implementation Details}\label{sec:imp_de}
Here, we discuss how to select the nuisance parameters, i.e., the interpolation parameter $\delta$ and the number of knots $N_{n_0}$, in the proposed ZIQSI method. 
As shown in the proof of Theorem \ref{thm:convergence}, a larger  $\delta$ is preferred for a faster convergence rate of the interpolation region, yet it may lead to a large variance at the change point. 
Note that our primary focus is constructing entire quantile curves rather than predicting conditional quantiles at a single $\tau$. Estimating the entire curve is generally insensitive to the choice of $\delta$, and we recommend $\delta = 0.499$ for simplicity. For prediction at a specific $\tau$, cross-validation can optimize $\delta$ for better performance \citep{ling2022statistical}.

To estimate $\beta_{\tau_s}$, which is required for estimating the quantile curve, we use equally spaced knots for the order $m$ B-spline with $N_{n_0} = \lfloor C n_0^{1/(2m+1)}\rfloor+1$, where $\lfloor a\rfloor$ denotes the integer part of a number, $C>0$ is a constant, and $n_0$ is the number of positive outcomes. The choice of $C$ does not change the estimation much in a reasonable range \citep{ma2016optimization}. In our numerical studies, we set $C = 1$ and choose $N_{n_0}$ by finding the first local  minimum of the following BIC criterion:
${\rm BIC}(N_{n_0}) = \log\left(L^{**}_{\hat\tau_s, n}(\theta)\right)+\frac{\log(n_0)}{2n_0}\left(N_{n_0}+m\right).$

\section{Simulations}\label{sec:simulation}
We present numerical experiments to assess the performance of the proposed ZIQSI, the method of \cite{ling2022statistical} (denoted as ``ZIQ-linear"), and the method of \cite{ma2016optimization} (denoted as ``Quantile Single-index"). We mainly focus on quantile-regression-based methods because \cite{ling2022statistical} already showed the superiority of their method compared to methods that require specific parametric assumptions (e.g., ZIP and ZINB), classic linear quantile regression without two-part modeling, and the Hurdle regression model.
ZIQ-linear can be viewed as a special case of ZIQSI by setting the function $G_{\tau_s}(\cdot)$ as an identity link function. Quantile Single-index performs similarly to the positive part of ZIQSI without adjusting $\tau$ by taking into account logistic regression. The Quantile Single-index model assumes the outcome to be continuous, and its estimation algorithm often fails to converge when the data contains a probability mass at zero. To circumvent this numerical difficulty, we added a small perturbation ($N(0,10^{-10})$) to the zero-valued outcomes and applied their method to the perturbed data.  For a fair comparison to ZIQ-linear, we use $\delta = 0.499$ for ZIQSI. Additional simulation results suggest that using a more minor $\delta$, such as $\delta = 0.250$, does not cause a significant difference in estimation (see Supplement S2.5).

Though ZIQSI provides estimates for both linear index $\beta_{\tau_s}$ and the function $G_{\tau_s}(\cdot)$, namely $\hat\beta_{\hat\tau_s}$ and $\widehat G_{\hat\tau_s}(\cdot)$, they are subject-specific and not comparable, as $\hat\tau_s = \Gamma(\tau;\textbf{x}, \hat\gamma)$ is a function of $\textbf{x}$. Therefore, we estimate quantile functions for $12$ individuals, whose health-related covariates $\textbf{x}$ are representative in real data (Table S2.1 in Supplement S2.1).  

To compare the performance of the three methods, we assess the estimated quantile curves $\widehat Q_Y(\tau\mid \textbf{x})$ by the relatively integrated mean squared error (RIMSE), the relatively integrated bias-squared (RIBIAS), and the relatively integrated variance (RIVAR) defined as follows: 
\[(1) 
 {\rm RIMSE} = \int \mathbb{E}\left\{\widehat{Q}_Y(\tau\mid \textbf{x}) - Q_Y(\tau\mid \textbf{x})\right\}^2 d\tau\bigg/\int Q_Y(\tau\mid \textbf{x})^2 d\tau,\]
\[(2){\rm RIBIAS} = \int \left\{\mathbb{E}\widehat{Q}_Y(\tau\mid \textbf{x}) - Q_Y(\tau\mid \textbf{x})\right\}^2 d\tau\bigg/\int Q_Y(\tau\mid \textbf{x})^2 d\tau,\] 
\[(3){\rm RIVAR} = \int \mathbb{E}\left\{\widehat{Q}_Y(\tau\mid \textbf{x}) - \mathbb{E}\widehat Q_Y(\tau\mid \textbf{x})\right\}^2 d\tau\bigg/\int Q_Y(\tau\mid \textbf{x})^2 d\tau.
\]
All three measurements are based on fixed $\textbf{x}$ and standardized by the squared scale of the quantile curve integrated through the entire process $\tau\in(0,1)$. The integrals in the three measurements are numerically approximated by the Riemann sums on $\tau = 0.01,0.02,\cdots,0.99$.

\subsection{Simulation settings}\label{sec:sim_set}
The dataset is simulated to mimic the real microbiome count data, with $Y$ being the read counts and
$\textbf{x}= (x_{1},x_{2},x_{3},x_{4},x_{5})^\top$ being covariates, according to the distribution from real data. For the covariates $\textbf{x}$, we generate $x_{1}\sim Bernoulli(0.5)$ for medicament, $x_{2}\sim N(28,2^2)$ for BMI, {$x_3\sim N(92.5,13^2)$ for waist circumference, $x_{4}\sim N(80,12^2)$ for diastolic blood pressure, and $x_5 \sim N (124, 18.5^2)$ for systolic blood pressure.}
For each dataset, we generated  $(\textbf{x}_i,y_i)$ for $i = 1,\cdots,n$ with the sample size $n=500$, similar to the sample size of the real data application. For the $i$th subject $\textbf{x}_i = (x_{i,1},\cdots,x_{i,5})^\top$, we first randomly simulated a variable {$\tau_i\sim Unif(0,1)$ representing the quantile level of the $i$th individual} and $D_i$ from a $Bernoulli$ distribution with a success probability defined as
$P(D_i = 1\mid \textbf{x}_i) = \pi(\gamma,\textbf{x}_i) = \frac{\exp(\gamma_0+\sum_{j=1}^5\gamma_{j}x_{i,j})}{1+\exp(\gamma_0+\sum_{j=1}^5\gamma_{j}x_{i,j})},$
where the parameter $\gamma = (-0.4,-0.480,-0.022,0.021,0.015, -0.009)^\top$ were set to control the proportion of zeros in outcomes. Then, we set $y_i = 0$ if $D_i=0$. If $D_i=1$, we generated the microbial count from the following quantile function: $ Q_Y(\tau_i\mid\textbf{x}_i,Y_i>0) = G_{\tau_i}\left(\beta_0(\tau_i)+\textbf{x}_i^\top\beta(\tau_i)\right),$
where the two sets of the true coefficients $\beta(\tau) = \left(\beta_1(\tau),\beta_2(\tau),\beta_3(\tau),\beta_4(\tau),\beta_5(\tau)\right)^\top$ and the quantile functions $G_{\tau}(\cdot)$ are simulated to mimic the distributions of a taxon in our real data analysis (see Supplement S2.1):
$\beta_0(\tau) = -147.7\tau- 50\tau^2-20,$ $\beta_{1}(\tau) = 0.6\sqrt{\tau}-2\tau$, $ 
\beta_{2}(\tau) = 2.2\tau^2$, $\beta_3(\tau) =  
\frac{2}{3}\tau^2-\frac{1}{3}\tau+0.4$, $\beta_{4}(\tau) = -0.1\sin(2\pi 
\tau)$, $\beta_5(\tau) = -0.6\tau^2+2\tau$, and $G_{\tau}(x) = \frac{1}{6}\tau x^4\times 
10^{-5}+\frac{1}{15}\tau x^2.
$
We provide the comparison between the distributions of the read count generated by our simulation setting and the read count of one real taxon count in Supplement S2.1 (Figure S2.1). Simulation results are presented based on 500 Monte Carlo replicates. Our method takes approximately 30 seconds to estimate the quantile regression model on a grid of nominal levels $\tau = 0.01,\cdots,0.99$ on a macOS machine with an Apple M2 chip.

\subsection{Results for model fitting}\label{sec:fit_re}
We use the samples $(\textbf{x}_i,y_i)$ for $i = 1,\cdots,500$ and the measurements  above. From Table \ref{sim:delta0.499}, we observe that the proposed ZIQSI method has a significantly smaller bias (RIBIAS) compared to both ZIQ-linear and Quantile Single-index. The RIVAR and RIMSE of ZIQSI and Quantile Single-index are comparable, and ZIQ-linear could have surprisingly large RIMSE due to large RIBIAS (e.g., subjects 1 and 2) as well as RIVAR (e.g., subjects 9 and 10). In general, ZIQ-linear performs worse than Quantile Single-index because the linear assumption on the quantile function for the positive part $Y>0$ is violated. {We also assess the simulation results where $G_\tau(x) = \tau x$ is a simple linear function in Supplement S2.7, and the result is consistent with our expectations.}
\spacingset{1}
 \begin{table}[!ht]
\caption{Summary of RIMSE($\%$), RIBIAS($\%$), RIVAR($\%$) of the estimated conditional quantile functions by ZIQSI, ZIQ-linear(ZIQ), and Quantile Single-index(QSI).}\label{sim:delta0.499}
\begin{tabular}{c|rrr|rrr|rrr}
\hline
& \multicolumn{3}{c|}{RIBIAS}                            & \multicolumn{3}{c|}{RIVAR}                              & \multicolumn{3}{c}{RIMSE}      
                     \\ \hline
ID           & ZIQSI &ZIQ   & QSI  & ZIQSI &ZIQ   & QSI  & ZIQSI &ZIQ   & QSI \\
\hline
1 & 0.19  & 21.18  &  1.14   & 3.20  & 5.25  & 2.81 & 3.39  & 26.43 &  3.95\\
2                    
&  0.07  & 21.29 & 0.44
& 3.96  & 6.19  &  3.94
& 4.03  &  27.48 & 4.38\\
 
3                   
& 0.24  & 4.07 & 1.10 
& 1.54  & 1.63  & 1.49 
& 1.78   & 5.70 & 2.59 \\
 
4                   
& 0.04  & 4.17  & 0.13 
& 1.67  & 1.66   & 1.82 
& 1.71  & 5.83 & 1.95 \\
 
5                  
& 0.10   & 2.53  &  0.76
& 3.31  &  1.01 & 3.62
& 3.41  & 3.54  &  4.38
\\
  
6                    
& 0.04  & 2.34  &  0.13
&  3.80 &  1.20 &  3.80
&  3.84  & 3.54  &  3.93
\\
   
7                    
& 0.34  & 1.23  &  0.84
&  3.09 &  2.00 &  2.95
&  3.43  & 3.23  &   3.79
\\
  
8                    
&  0.12  & 1.27  &  0.65
&  3.54 &  2.27 &  3.55
& 3.66  & 3.54  &  4.20
\\
   
9                   
&  0.13  & 19.12  & 1.01 
& 1.57  &  4.30 &  2.36
& 2.70  & 23.42  &  3.37
\\
 
10                   
& 0.02  & 18.88  &  0.15
& 3.01  & 4.83  &  3.14
& 3.03  & 23.71  &  3.29
\\
   
11                 
&  0.02  & 9.04  &  0.99
& 1.98  &  2.22 &  1.53
& 2.00  & 11.26  &  2.54
\\
  
12                  
& 6.26$e^{-5}$  & 9.93  &  0.12
&  2.25 &  2.55 & 2.42
&  2.25 & 12.48  & 2.54  
\\
 \hline
\end{tabular}
\end{table}
\spacingset{2}

We further report the average proportion of negative predicted counts over the quantile process $\tau\in(0,1)$ for three methods in Table S2.2 in Supplement S2.2. Based on Table \ref{sim:delta0.499} and Table S2.2 in Supplement S2.2, we observe that ZIQ-linear has a small portion of negative predictions but severe bias; meanwhile, Quantile Single-index suffers from a large portion of negative predictions, though the estimation bias is moderate. {To eliminate the effect of results below zero, we present the results truncated at zero in Supplement S2.3 (see Table S2.3), where our method remains its advantages.} The proposed method ZIQSI shows its superiority regarding the smallest integrated bias and a reasonably small portion of predictions below zero. 

For each subject, we also visualize the estimation performance of each method across the quantile process $\tau\in(0,1)$. We reported the average estimated quantile curves and their 95\% confidence intervals based on the $500$ estimations above. The confidence interval is constructed based on the percentile of the empirical distribution of  $\widehat{Q}_Y(\tau\mid \textbf{x})$ at a given $\tau$. We show subject 11 in Figure \ref{fig:id11} and present others in Supplement S2.4. We observe that both Quantile Single-index and ZIQ-linear have an obvious bias, and a larger estimation bias of ZIQ-linear is observed at upper quantiles. Similar patterns are observed for other individuals (Supplement S2.4). {The simulation results for AQE indicate that ZIQSI provides the most accurate and stable estimation compared to the other two methods (Supplement S2.6).}

\begin{figure}[t!]
\centering
\includegraphics[scale = 0.43]{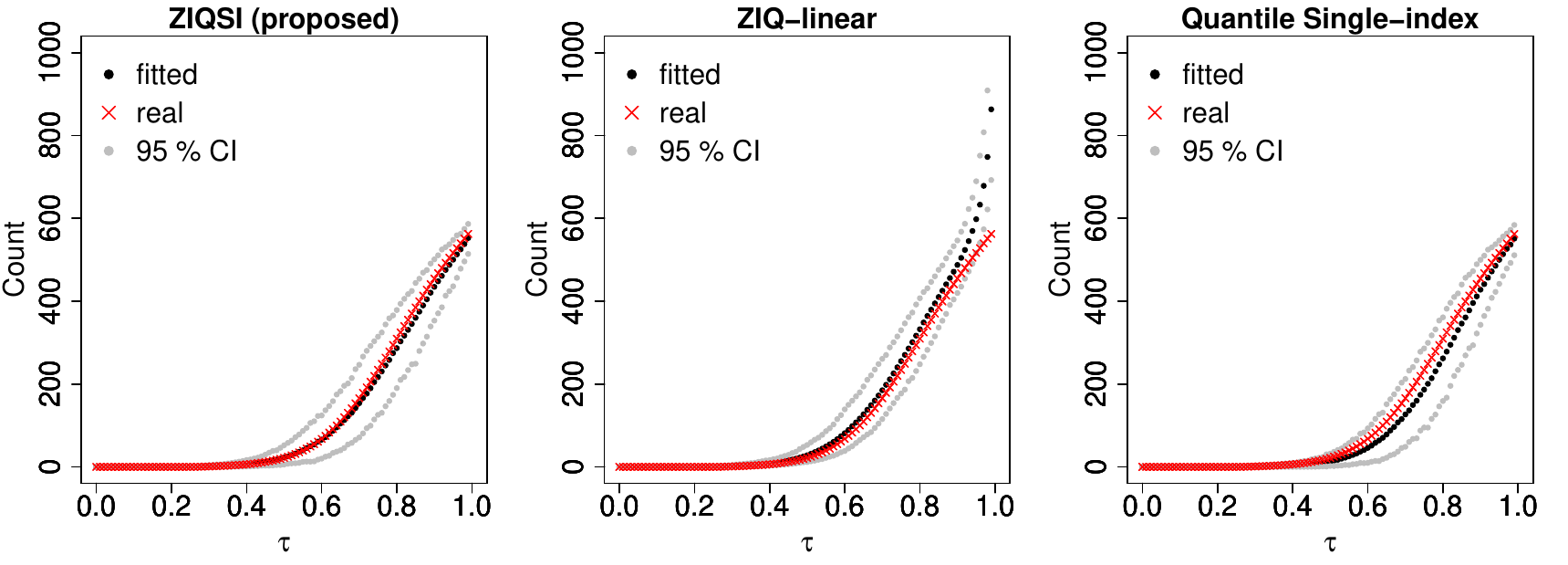}
\caption{Quantile curves based on $500$ times estimations (Subject $11$).}
\label{fig:id11}
\end{figure}

\section{Application}\label{sec:app}

 In this section, we illustrate the performance of our ZIQSI method by the study of Columbian's Gut \citep{de2018gut,qiita}. 
We compare the proposed method (ZIQSI) with the method of \cite{ling2022statistical} (ZIQ-linear) and the method of \cite{ma2016optimization} (Quantile Single-index) by assessing model fitting from the population and individual perspectives.  As in Section \ref{sec:simulation}, we use $\delta = 0.499$ for a fair comparison with ZIQ-linear. The results of using a smaller $\delta$ are similar and we provided them in Supplement S3.1. 

\subsection{Data description} 

\begin{figure}[!ht]
\centering
\begin{subfigure}{0.45\textwidth}
  \includegraphics[scale = 0.3]{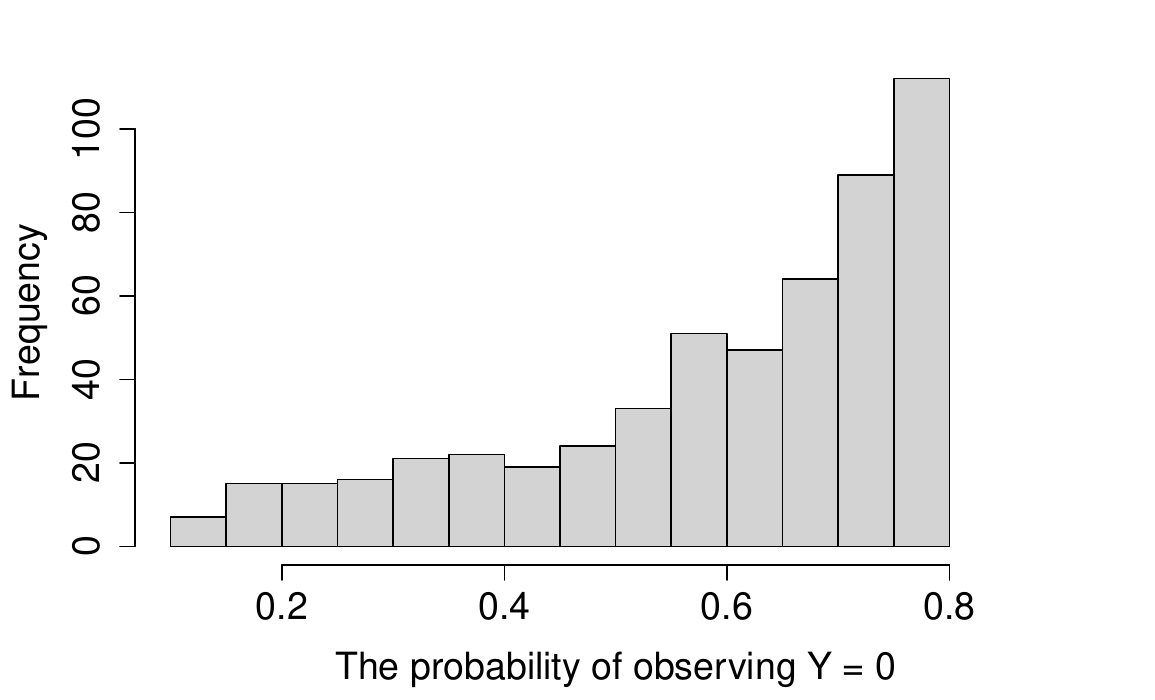} 
  \caption{Zero proportion per taxon.}
\end{subfigure}
\begin{subfigure}{0.45\textwidth}
\includegraphics[scale = 0.3]{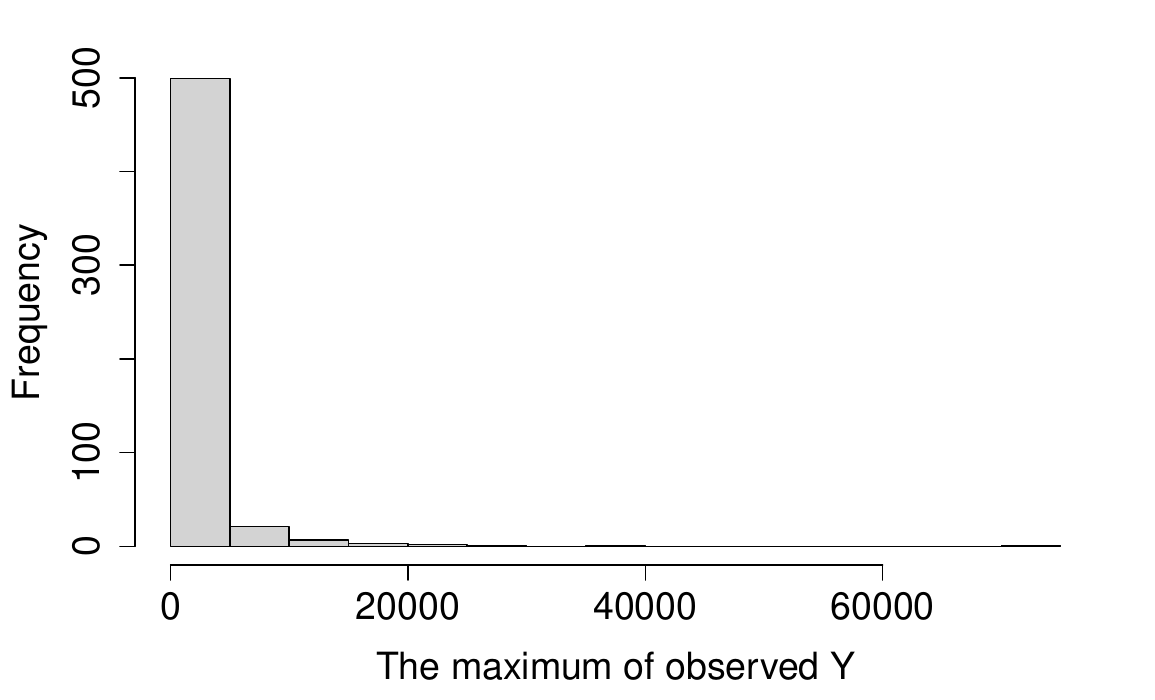}
 \caption{Maximum count per taxon.}
\end{subfigure}
\caption{Histogram for microbiome counts.}\label{fig:hist}
\end{figure}

The dataset contains microbiome counts of over $6000$ taxa for $441$ adults, along with covariates related to diet, obesity, and cardiometabolic diseases. We consider taxa with observed zero proportions less than $0.8$, as a larger percentage of zeros commonly leads to unreliable results \citep{wadsworth2017integrative,jiang2021bayesian,zhang2020fast}. From Figure \ref{fig:hist}, we observe that a large number of taxa are heavily zero-inflated, and the observed counts are overdispersed.

Following the study of \cite{de2018gut}, we analyzed taxa counts with health-related covariates as follows: anthropometric measures (age, BMI, sex, waist circumstance), lipid profile (adiponectin, total cholesterol,  HDL, LDL, triglycerides), glucose metabolism (glucose, glycosylated hemoglobin, insulin), blood pressure (diastolic blood pressure, systolic blood pressure), city,  medicament, and macronutrient consumption (fiber, percentage of animal protein, carbohydrates, monounsaturated fat, polyunsaturated fat, saturated fat, total fat, protein). Among them, categorical variables, such as sex, medicament, and city, are treated as dummy variables. We further removed $3$ subjects for missing values and $2$ subjects for extremely high values of triglycerides over $800$ $mg/dL$, resulting in $436$ samples in our analysis. We analyzed $535$ taxa with observed zero proportions in the range of $0.1$-$0.8$. 
As a common practice in other microbiome studies \citep{xia2018modeling}, we adjust for the library size, which is the sum of all $535$ taxa counts per person.

\subsection{Goodness-of-fit}\label{sec:app_good}

To provide a thorough analysis, we used a representative taxon, namely $Slackia$, which has the third highest abundance out of 97 taxa in the $Coriobacteriaceae$ family from the co-abundance groups $Prevotella$ based on hierarchical clustering with Ward's linkage \citep{claesson2012gut}. We assessed model fitting and quantile curve estimation from different perspectives. We also analyzed taxa with varying degrees of zeros and provided results in Supplement S3.2. 

To assess the goodness of fit for a model, we adopt the measurement used in \cite{ling2022statistical} and \cite{heyman1991statistical} to compare the distribution of the observed data and the predicted values from fitted models. We first fit the model based on the aforementioned three methods. Then, a quantile level $\tau$ is randomly drawn from $ Unif(0,1)$, and $\widehat Q_Y(\tau\mid X)$ is reported as the fitted microbiome counts given observed covariates. {The computation time for estimating the quantile single-index models on the nominal quantile levels $\tau = 0.01,\cdots,0.99$ is around 32 seconds on a macOS machine with an Apple M2 chip.}

\begin{figure}[!ht]
    \centering  \includegraphics[scale = 0.4]{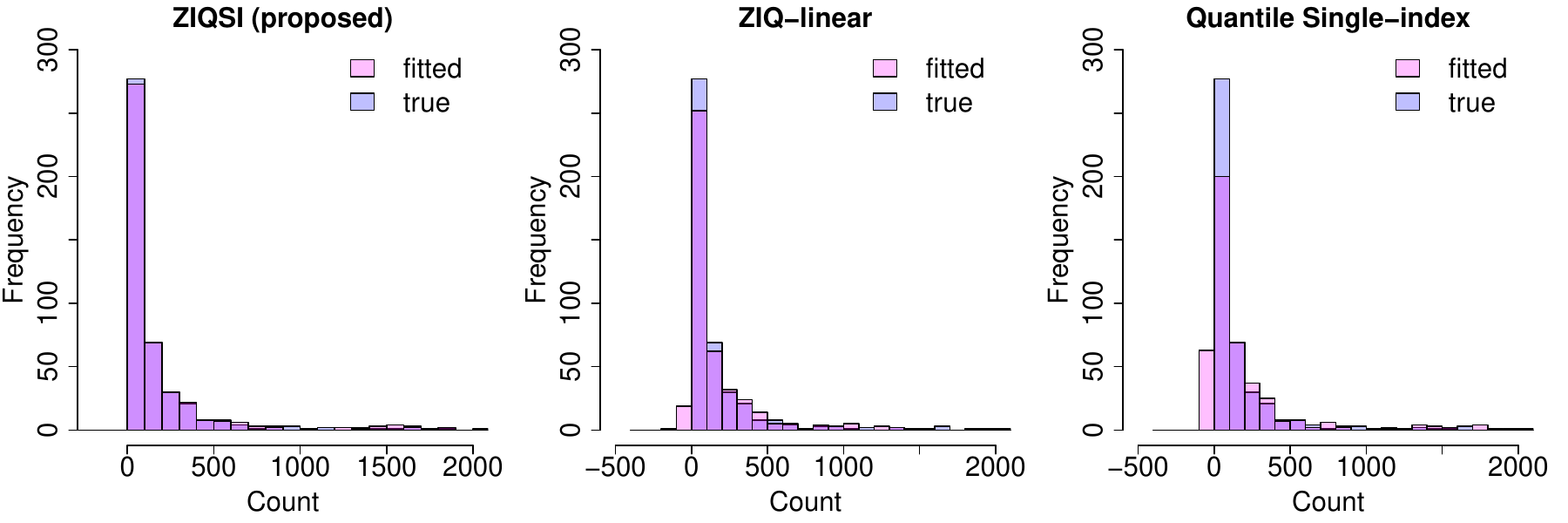}
 \caption{Histogram plot of $Slackia$.}\label{fig:hist_12}
\end{figure} 

From Figure \ref{fig:hist_12}, we observe that the proposed ZIQSI method better fits the taxon $Slackia$ compared to the other two methods, especially at two tails. 
On the contrary, both  ZIQ-linear and Quantile Single-index predicted counts below zero, which is against the non-negativity of the number of microorganisms.  Model fitting results for other taxa also suggest a similar pattern (see Supplement S3.2). Though ZIQ-linear commonly has a smaller proportion of negative predicted counts compared to Quantile Single-index, the values could be as small as $-500$ (Supplement S3.2 Figure S3.8). 
Quantile Single-index often has many negative predicted counts, which is consistent with the simulation results.
To investigate the lack of goodness of fit for ZIQ-linear and Quantile Single-index methods, we provide detailed discussions from the population and individual perspectives below.

\subsection{Estimated quantile curves}\label{sec:qrcurve}
As the quantile effect is caused by the logistic and quantile single-index components, visualizing it is more complicated than simply presenting $\hat\beta_\tau$ or $\widehat G_\tau(\textbf{x}^\top \hat\beta_\tau)$. It needs to be highlighted that the effect of covariates in the logistic regression also plays a role through $\Gamma(\tau;\textbf{x}, \hat\gamma_n)$ (i.e., $\hat\tau_s$). That is, given a fixed $\tau$ and $\hat\gamma_n$ estimated from logistic regression, $\hat\beta_{\hat\tau_s} = \hat\beta_{\Gamma(\tau;\textbf{x}, \hat\gamma_n)}$ is a function of $\textbf{x}$. Thus, we visualize the quantile effects for covariates by fixing $\tau$ while changing $\textbf{x}$, or vice versa. 

\begin{figure}[!ht]
\centering
\includegraphics[scale = 0.4]{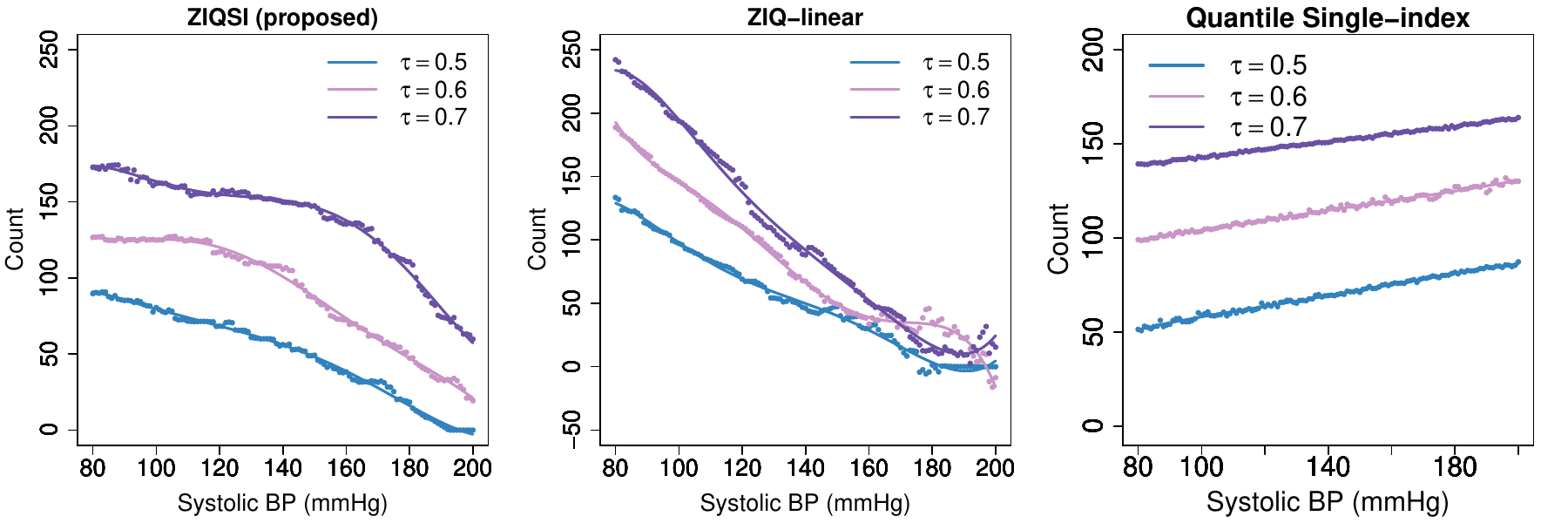}
\caption{Predicted counts for taxon  $Slackia$ with the change of systolic bp (other covariates are fixed). }
\label{fig:systolic-bp}
\end{figure}

First, we present how $\widehat{Q}_Y(\tau;\textbf{x})$ changes with $\textbf{x}$ at given $\tau$.
For illustration, we consider the distinct variable systolic blood pressure (denoted as ``systolic bp") as the target covariate and take the other covariates fixed, since systolic bp has a significant effect on the abundance of $Slackia$ \citep{de2018gut}. Specifically, the continuous covariates are fixed at their average levels, and we take binary/categorical covariates ``sex" as female, ``medicament" as $1$, and ``city" as Cali. Using the microbiota $Slackia$ as an example again, we present the estimated quantile curves $\widehat G_\tau(\textbf{x}^\top\beta(\tau))$ regarding different levels of systolic bp at the nominal quantile levels $\tau = \{0.5, 0.6, 0.7\}$ in Figure \ref{fig:systolic-bp}. Of note, though the nominal quantile level $\tau$ is fixed, $\tau_s$, adjusted by the logistic regression, changes with different levels of systolic bp. Thus, the points in Figure \ref{fig:systolic-bp} do not align well, and we provided  B-spline fitted curves based on the estimated points. We observe that ZIQ-linear has the quantile crossing issue when the systolic bp is larger than 160. That is, the estimated counts at $\tau = 0.6$ are lower than the ones at $\tau = 0.5$ and higher than the ones at $\tau=0.7$, which violates the monotonic nature of quantiles. Also, the predicted counts at $\tau=0.6$ with ZIQ-linear are negative with large systolic bp values, which explains the negative predicted values we observed in the histogram (Figure \ref{fig:hist_12}). 

\begin{figure}[!ht]%
\centering\begin{subfigure}{0.3\textwidth}\centering\includegraphics[scale = 0.33]{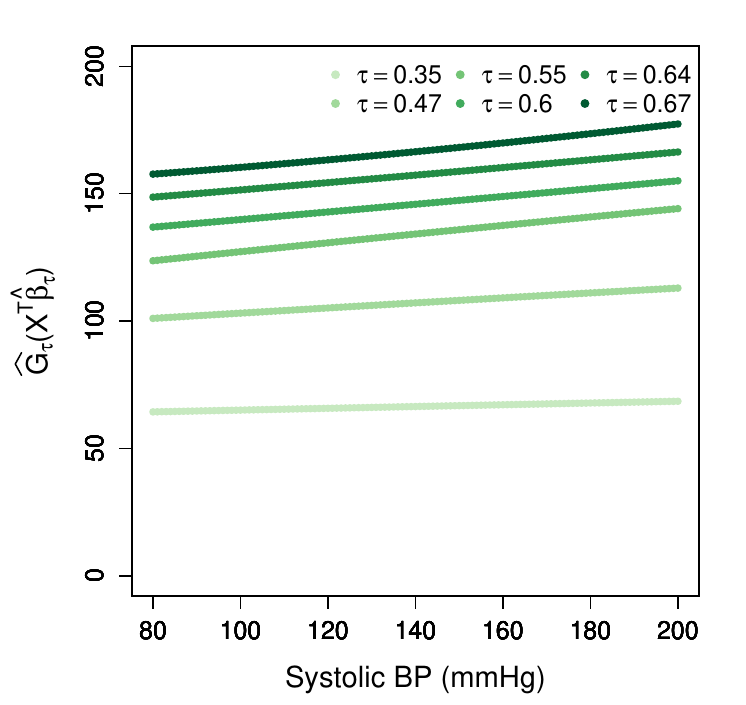}\caption{}\end{subfigure}\begin{subfigure}{0.34\textwidth}\centering\includegraphics[scale = 0.33]{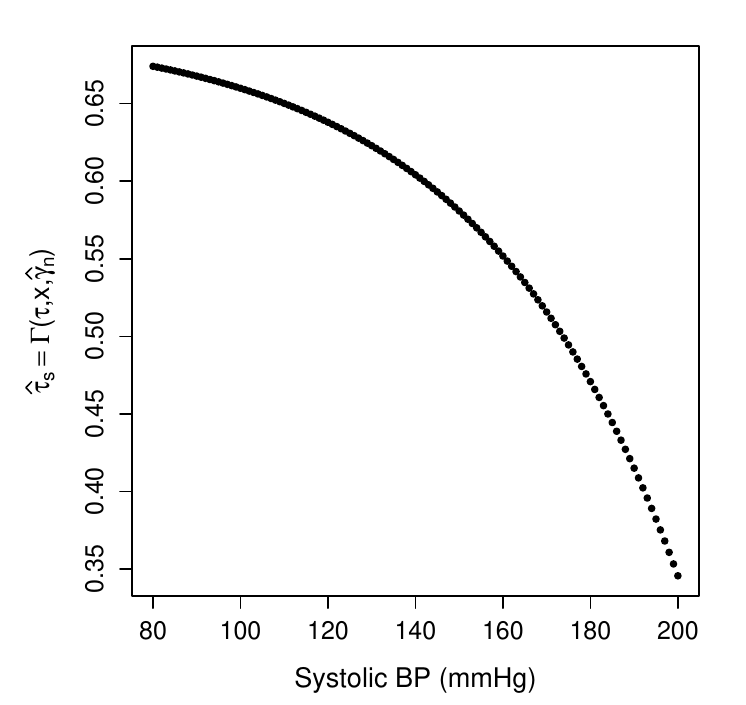}\caption{ }\end{subfigure}\begin{subfigure}{0.3\textwidth}\centering\includegraphics[scale = 0.33]{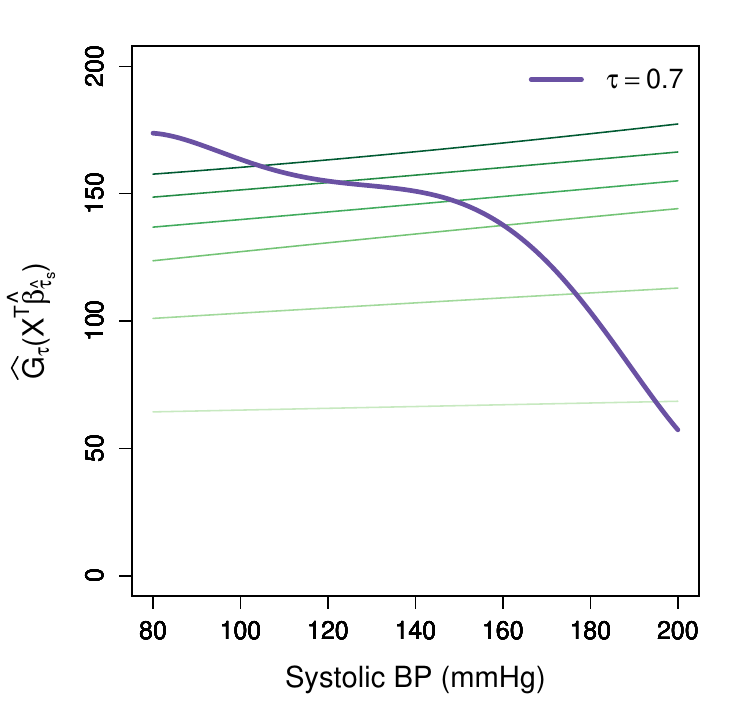}\caption{}\end{subfigure}\caption{Compare the estimated quantile curve with original and adjusted quantile levels. (a): Estimated quantile curves from ZIQSI with unadjusted $\tau$. (b): Mapping $\tau=0.7$ to $\hat\tau_s$ through $\Gamma(\tau,\textbf{x},\hat\gamma_n)$ with the change of systolic bp ($\textbf{x}$). (c):  Estimated quantile curves from ZIQSI with adjusted $\hat\tau_s$ (purple curve), while the quantile level is fixed at $\tau = 0.7$.}\label{fig:unadjusted}
\end{figure}

From Figure \ref{fig:systolic-bp}, we observed that the predicted counts from ZIQ-linear and ZIQSI have a similar decreasing pattern with the increase of systolic bp, though ZIQ-linear has some unreasonable predictions. The estimated quantile curves by the Quantile Single-index method, however, showed a different trend as it does not adjust the quantile level $\tau$ and assumes the probability of observing a zero outcome is the same for every subject. Thus, we further illustrate the difference between the pre-fixed quantile level $\tau$ and its adjusted version $\hat\tau_s$. In Figure \ref{fig:unadjusted}(a), the quantile curves estimated by ZIQSI with the fixed quantile levels $\tau$ have similar trends as the curves estimated by the Quantile Single-index method (Figure \ref{fig:systolic-bp} (right)). Then, when we consider a quantile level $\tau = 0.7$, its mapped quantile level $\hat\tau_s$ is decreasing with the increase of systolic bp owing to its negative effect (Figure \ref{fig:unadjusted}(b)), as systolic bp has a negative estimated coefficient in the logistic regression, which means a higher systolic bp level can lead to a lower nominal $\hat\tau_s$. Naturally, the change of $\hat\tau_s$ results in the accelerated decreasing curve $\widehat{G}_{\hat\tau_s}(\textbf{x}^\top \hat\beta_{\hat{\tau}_s})$ (Figure \ref{fig:unadjusted}(c)), which is consistent with the results presented in Figure \ref{fig:systolic-bp} (left). For ZIQ-linear, the trend of its estimated quantile curves is similar to ZIQSI, as the adjustment for $\tau$ through logistic regression (i.e., $\hat\gamma_n$) remains the same. 

\begin{figure}[!ht]
\centering
\includegraphics[scale = 0.4]{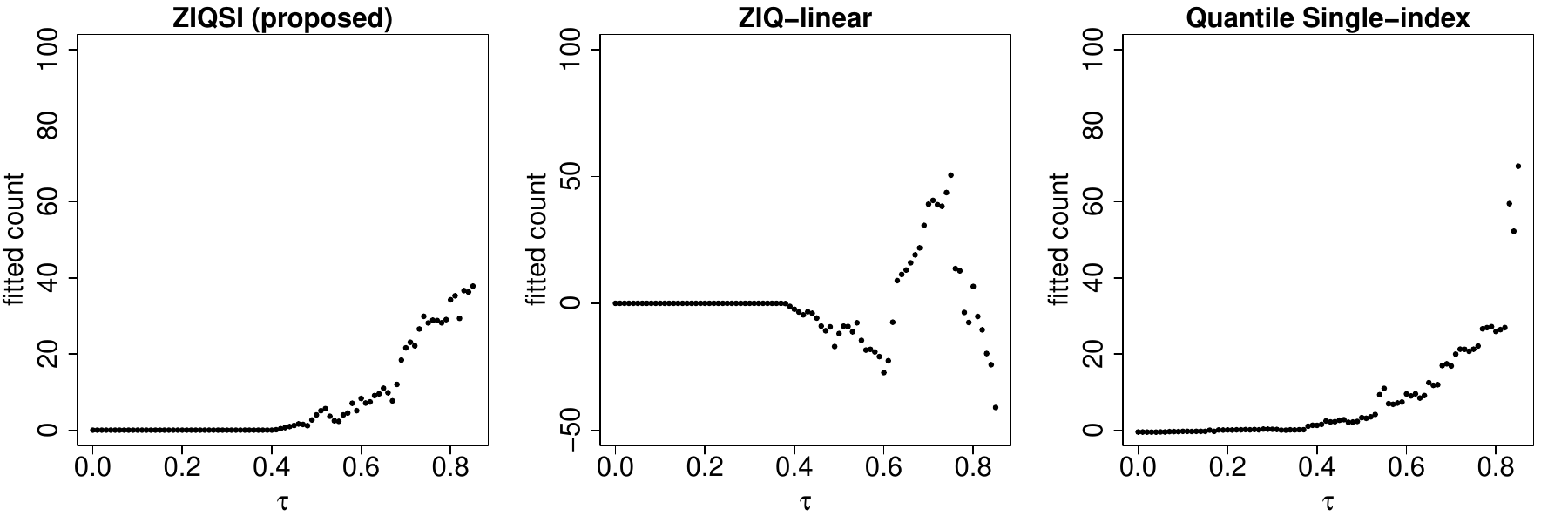}
\caption{Predicted quantile curve of subject X11993.MI385H.}
\label{fig:systolic-bp2}
\end{figure}

Then, we show the estimated quantile curve $\widehat{Q}_Y(\tau;\textbf{x})$ for a specific subject. Among the subjects whose predicted counts are negative, we randomly select one sample and present the fitted quantile curves (Figure \ref{fig:systolic-bp2}). The proposed ZIQSI method reasonably estimates the entire quantile curve, while ZIQ-linear showed a non-monotone curve with the increase of $\tau$, which is counter-intuitive and against the nature of quantiles. Further,  ZIQ-linear has negative predictions, which is counter-intuitive as the response is required to be non-negative. We also present the effect of a specific covariate by comparing the AQE based on each method (see Supplement S3.3).

\section{Discussion}\label{sec:discussion}
In this paper, we focus on statistical modeling for zero-inflated and overdispersed microbiome data. To relax parametric assumptions in existing two-part modeling approaches and provide more flexibility in handling complex associations, we propose a novel semiparametric single-index quantile regression model that first extends single-index quantile regression models to zero-inflated and overdispersed outcomes. Both the theoretical and empirical works suggest that this method outperformed in modeling zero-inflated and overdispersed outcomes. 

Several interesting topics warrant further investigation. First, current quantile regression methods for zero-inflated data, including our ZIQSI method, do not enforce non-negativity, potentially leading to negative predictions due to numerical issues. Adding a non-negativity constraint to the link function $G_\tau$  could address this; see \citep{cannon2018non} for an example in composite quantile regression.  Next, while our method accommodates high-dimensional covariates via single-index models, it may struggle with high-dimensional data. Incorporating regularization \citep{li2008sliced, peng2011penalized} or using semiparametric dimension reduction \citep{ma2012semiparametric} could improve performance, though this would require refining the asymptotic theory. Challenges also arise when the number of covariates grows with sample size, complicating the nonparametric estimation of  $G_\tau(\cdot)$  and the linear index. Lastly, our method focuses on normal quantile levels, but estimating tail quantiles is particularly challenging, especially as $\tau_n$ approaches 1 at the rate $n(1-\tau_n) \to c > 0$ \citep{xu2022extreme}. Extending the tail single-index model \citep{xu2022extreme} to zero-inflated data presents a promising avenue, as zero inflation further complicates extreme quantile estimation by reducing the effective sample size.

\paragraph{Data and Software Availability:}
The data used in Section \ref{sec:app} is publicly available at https://qiita.ucsd.edu/.  We have developed the \texttt{R} package and published it on Github: https://github.com/tianyingw/ZIQSI/.

\section*{Supplementary Materials:}
The supplement contains the proofs of the theorems and the additional results for simulation and application.
\section*{Acknowledgements}
We thank the editor, associate editor, and two referees for their valuable 
comments and constructive suggestions.\par


\bibhang=1.7pc
\bibsep=2pt
\fontsize{9}{14pt plus.8pt minus .6pt}\selectfont
\renewcommand\bibname{\large \bf References}
\expandafter\ifx\csname
natexlab\endcsname\relax\def\natexlab#1{#1}\fi
\expandafter\ifx\csname url\endcsname\relax
\def\url#1{\texttt{#1}}\fi
\expandafter\ifx\csname urlprefix\endcsname\relax\def\urlprefix{URL}\fi
\bibliographystyle{chicago}      
\bibliography{SDR_tw.bib}   

\vskip .65cm
\noindent
Department of Statistics and Data Science, Tsinghua University
\vskip 2pt
\noindent
E-mail: wzr23@mails.tsinghua.edu.cn
\vskip 2pt

\noindent
Department of Statistics, Colorado State University
\vskip 2pt
\noindent
E-mail: Tianying.Wang@colostate.edu

\end{document}